# Taxonomy-aware distances between scholarly topic profiles via an exact simplex embedding

Dmitry Gubanov and Alexander Chkhartishvili

**Abstract**

Topic profiles represent publications, authors, and other scholarly entities as probability distributions over a fixed set of topics, but flat total variation treats every pair of distinct pure-topic profiles as maximally separated and therefore ignores taxonomic proximity. From a rooted weighted taxonomy, we derive a cardinality-normalized linear operator that maps the leaf topics to points in the original probability simplex and exactly realizes a normalized lowest-common-ancestor ultrametric under total variation. The operator is doubly stochastic and positive definite; within the class of nonnegative edge-cluster Gram operators, its normalization is uniquely determined on the reduced branching tree. Applying the same invertible operator to arbitrary topic mixtures yields a nondegenerate hierarchy-aware metric that contracts flat total variation, differs from the tree-Wasserstein distance on mixtures, and can be evaluated in $O(|V| + L)$ time and memory without forming the dense matrix. In a frozen OpenAlex taxonomy with 4,516 terminal Topics, raw dissimilarities between Topic texts showed consistent ordinal alignment with taxonomic proximity, while only 3 of 253 calibrated internal nodes required monotonic correction. Encoder choice nevertheless affected individual height estimates. The framework exactly realizes a supplied weighted hierarchy; text is used only to initialize its node heights, and distances between scholarly topic profiles are then computed in the induced geometry.

## 1 Introduction

Topical classification underpins many forms of scientometric analysis. Most established classification systems assign publications to one or more research fields. In traditional journal-level schemes, journals are classified in advance and their articles inherit that classification (Clarivate, n.d.), although publication-level alternatives are increasingly being developed (Waltman and van Eck, 2012; Álvarez-Llorente et al., 2025).

A common downstream representation is a topic profile: a probability vector whose coordinates quantify the association of a publication, author, organization, journal, or conference with a fixed set of topics (Schaefermeier et al., 2021). Such profiles support retrieval, portfolio analysis, and science mapping. However, under flat total variation, every pair of distinct one-hot topic profiles is maximally separated, irrespective of the taxonomy. This geometry is difficult to justify when the coordinates are leaves of a scientific knowledge hierarchy.

Distance design for scholarly profiles is an informetric measurement problem rather than merely a choice of vector norm. Previous work has compared Euclidean, barycentric, cosine-based, and similarity-adapted distances between publication portfolios (Rousseau et al., 2017). Other studies have derived distances between disciplines from collaboration patterns and evaluated similarity-to-distance transformations (Vancraeynest et al., 2024), still others have constructed patent-level exploration distances from citation-network embeddings (Choi and Yoon, 2022).

Related studies generalize citation-based similarity (Yun, 2022), fuse multiple journal-similarity layers (Baccini et al., 2022), or evaluate hierarchical and semantic representations used to construct topics and profiles (Urdiales and Guzmán, 2024; Chen et al., 2024). None of these approaches provides all of

D. Gubanov is with the Institute of Control Sciences, RAS, Moscow, Russia (dmitry.a.g@gmail.com).
A. Chkhartishvili is with the Institute of Control Sciences, RAS, Moscow, Russia (sandro_ch@mail.ru).

the properties required here: preservation of the probability simplex, exact recovery of a specified leaf ultrametric, injectivity on mixtures, and tree-linear distance evaluation.

We therefore seek a taxonomy-aware distance that preserves probability-valued profiles and the interpretation of total variation. The problem was motivated by the ISAND ontology for control theory and its applications (Gubanov et al., 2024), but the formulation below requires only a rooted taxonomy.

Tree metrics have classical $\ell_1$ and cut-metric representations (Deza and Laurent, 1997); hierarchy-induced similarities form a known class of positive-semidefinite kernels (Kriege et al., 2016); and tree-Wasserstein distances compare distributions by transporting mass along a tree (Le et al., 2019; Peyré and Cuturi, 2019). The closest methodological analogue is the similarity-adapted publication-vector approach, in which a publication-count vector is transformed with an externally supplied similarity matrix (Rousseau et al., 2017).

To satisfy these requirements jointly, we derive an exact probability-simplex realization of the normalized lowest-common-ancestor ultrametric. On the reduced branching tree, the cardinality normalization is unique within the nonnegative edge-cluster operator class; the resulting operator is doubly stochastic, positive definite, and invertible, and it induces a hierarchy-aware metric on mixed profiles with matrix-free evaluation in $O(|V|+L)$ time. A frozen OpenAlex case study covering 4,516 Topics then examines raw taxonomy–text ordering, the monotonic correction required to obtain admissible heights, numerical recovery of the theoretical identities, and sensitivity to the height-estimation statistic and encoder.

# 2 Scholarly topic profiles and the measurement problem

## 2.1 From ontology to a rooted taxonomy

Following Gruber (1993), an ontology is an explicit specification of a conceptualization. Unlike a taxonomy, it may contain multiple types of concepts and relations and need not form a tree.

This work was motivated by the ISAND ontology and its profile-construction framework (Gubanov et al., 2024); however, the analysis uses only the narrower structure required by the mathematics: a hierarchical taxonomy represented as a rooted tree. OpenAlex provides a current large-scale example, organizing scholarly works into an approximately 4,500-topic, four-level hierarchy of Domains, Fields, Subfields, and Topics (Priem et al., 2022; OpenAlex, n.d.). In the frozen OpenAlex snapshot analyzed in Section 4, terminal Topics serve as profile coordinates and higher taxonomic ranks as internal categories.

## 2.2 Profiles of publications and authors

Let the hierarchical taxonomy be represented by a rooted tree $T=(V,E)$ with leaf set

$$\mathcal{L}=\{\ell_1,\dots,\ell_L\},\qquad L=|\mathcal{L}|.$$

The following construction, inherited from the ISAND framework (Gubanov et al., 2024), is one possible estimator of a topic profile. The distance developed in Section 3 does not depend on this particular estimator. Use the following notation:

- $\mathcal{K}$ be the set of researchers, $K=|\mathcal{K}|$, and $k\in\{1,\dots,K\}$ a researcher index;
- $\mathcal{M}$ be the set of publications, $M=|\mathcal{M}|$, and $i\in\{1,\dots,M\}$ a publication index;
- $\Delta_{ij}$ be the total number of occurrences of terms associated with leaf topic $j$ in publication $i$, where $j\in\{1,\dots,L\}$;
- $\omega(k,i)$ be the authorship indicator

$$\omega(k,i)=\begin{cases}1, & \text{if researcher } k \text{ is an author of publication } i,\\ 0, & \text{otherwise;}\end{cases}$$

- $r(i)$ be the number of authors of publication $i$.

The *topic profile of publication* $i$ is

$$x_i = (x_{i1}, \dots, x_{iL}), \qquad x_{ij} = \frac{\Delta_{ij}}{\sum_{q=1}^{L} \Delta_{iq}}, \qquad j \in \{1, \dots, L\}. \tag{1}$$

Provided that $\sum_{q=1}^{L} \Delta_{iq} > 0$, the vector $x_i$ belongs to the standard probability simplex

$$\Delta^{L-1} = \left\{ x \in \mathbb{R}^L : x_j \geq 0, \ \sum_{j=1}^{L} x_j = 1 \right\}.$$

For publications in which no terms from the taxonomy are detected, Eq. (1) is undefined, so a separate fallback rule is required.

Publication profiles can in turn be aggregated to define profiles of other scholarly entities. For example, the topic profile of author $k$ can be defined from the author's publication record using fractional authorship weights:

$$y_{kj} = \frac{\sum_{i=1}^{M} \omega(k,i) \frac{x_{ij}}{r(i)}}{\sum_{q=1}^{L} \sum_{i=1}^{M} \omega(k,i) \frac{x_{iq}}{r(i)}}, \qquad j \in \{1, \dots, L\}. \tag{2}$$

This profile is defined when author $k$ has at least one publication with a valid profile and a positive authorship weight. Profiles of journals, conferences, and organizations can be defined analogously.

### 2.3 Flat-topic baseline

Scholarly entities represented by probability vectors can be regarded as points on a standard simplex. For example, reviewer matching can compare a manuscript profile with the profiles of candidate reviewers or with profiles of their publications.

For probability vectors $x = (x_1, \dots, x_L)$ and $y = (y_1, \dots, y_L)$, we adopt the total variation distance

$$d_{\mathrm{TV}}(x, y) = 1 - \sum_{j=1}^{L} \min(x_j, y_j) = \frac{1}{2} \sum_{j=1}^{L} |x_j - y_j|. \tag{3}$$

It coincides with one half of the Manhattan distance on the probability simplex. Its overlap form is interpretable: topics $\ell_j$ for which $\min(x_j, y_j)$ is comparatively large contribute most strongly to the overlap between the two profiles. Equivalently, total variation is the 1-Wasserstein distance for the discrete ground cost that assigns unit cost to every pair of distinct topics (Le et al., 2019; Peyré and Cuturi, 2019). The flat baseline therefore explicitly assumes that all distinct topics are equally distant.

Let $e_i$ denote the one-hot profile concentrated on topic $\ell_i$. Equation (3) gives

$$d_{\mathrm{TV}}(e_i, e_j) = 1 \qquad \text{for every } i \neq j.$$

Thus two sibling leaves and two leaves whose lowest common ancestor is the root are equally distant under the flat baseline. A taxonomy-aware geometry should distinguish these cases without abandoning probability-valued profiles.

Accordingly, the construction must satisfy four desiderata: (i) preserve probability-valued profiles with leaf-indexed coordinates; (ii) recover the prescribed distances between pure leaf topics exactly; (iii) remain injective when extended linearly to mixed profiles; and (iv) permit evaluation without storing a dense $L \times L$ matrix.

# 3 Taxonomy-aware topic-space construction

## 3.1 Formal setting

We now generalize the flat topic space by allowing a hierarchy to prescribe the distances between leaf topics. The points to be represented are the leaves of a rooted tree, and node heights determine their pairwise distances.

Let $T = (V, E)$ be a rooted tree with leaf set $\mathcal{L} = \{\ell_1, \dots, \ell_L\}$, where $L \geq 2$. Let $h : V \to \mathbb{R}$ satisfy:

1. *Height normalization:* $h(\text{root}) = 1$, $h(v) > 0$ for every internal node $v$, and $h(\ell) = 0$ for every leaf $\ell \in \mathcal{L}$.
2. *Strict height monotonicity:* for each edge $e = (p, c) \in E$ from parent $p$ to child $c$, $h(p) > h(c)$. The edge weight is $w_e = h(p) - h(c) > 0$.

For $i, j \in \{1, \dots, L\}$, define

$$H_{ij} = h\big(\text{lca}(\ell_i, \ell_j)\big), \tag{4}$$

where $\text{lca}(\ell_i, \ell_j)$ is the lowest common ancestor of leaves $\ell_i$ and $\ell_j$. The objective is to associate each leaf $\ell_k$ with a probability vector $b^{(k)}$ such that

$$d_{\text{TV}}\big(b^{(i)}, b^{(j)}\big) = H_{ij} \qquad \text{for all } i, j \in \{1, \dots, L\}. \tag{5}$$

Under these assumptions, $H$ is an ultrametric. Indeed, among the three pairwise lowest common ancestors of any three leaves, two coincide and are ancestors of the third; because heights decrease toward the leaves, the two largest pairwise values of $H$ are equal. Moreover, if each edge $e$ is assigned length $w_e/2$, telescoping along the two branches from $\text{lca}(\ell_i, \ell_j)$ shows that the weighted leaf-to-leaf path distance equals $H_{ij}$.

## 3.2 Simplex embedding algorithm

Index the leaves in a fixed order, for example, by a left-to-right traversal. The coordinates of $b^{(k)} = (b_1^{(k)}, \dots, b_L^{(k)}) \in \mathbb{R}^L$ are indexed in the same leaf order. For each edge $e = (p, c)$, let

$$S_e = \{\ell_j \in \mathcal{L} : \ell_j \text{ lies in the subtree rooted at } c\},$$
$$m_e = |S_e|, \qquad w_e = h(p) - h(c).$$

The following algorithm constructs the vectors directly.

**Algorithm 1: Embedding the leaves of a rooted taxonomy into the probability simplex.**
**Input:** a rooted tree $T = (V, E)$, its leaf set $\mathcal{L}$, and a height function $h$.

1. Initialize $b^{(k)} \leftarrow (0, \dots, 0) \in \mathbb{R}^L$ for every $k \in \{1, \dots, L\}$.
2. For every edge $e = (p, c) \in E$, calculate $S_e$, $m_e = |S_e|$, and $w_e = h(p) - h(c)$. For every $\ell_k \in S_e$ and every $\ell_j \in S_e$, update
$$b_j^{(k)} \leftarrow b_j^{(k)} + \frac{w_e}{m_e}.$$
3. Return $b^{(1)}, \dots, b^{(L)}$.

Because Algorithm 1 explicitly initializes and returns $L$ dense vectors of length $L$, its running time is

$$O\left(L^2 + \sum_{e \in E} |S_e|^2\right),$$

and its output requires $\Theta(L^2)$ storage. A general upper bound is $O(L^2 + |E|L^2)$. After contracting unary internal nodes, $|E| = O(L)$, giving an $O(L^3)$ worst-case bound; for a balanced reduced tree, the sum $\sum_e |S_e|^2$ is $O(L^2)$.

### 3.3 Operator form and relation to hierarchy-induced kernels

Let $u_e \in \{0,1\}^L$ be the indicator vector of the leaf set $S_e$, and let $B_h \in \mathbb{R}^{L\times L}$ denote the operator induced by the normalized height function $h$, with its $k$-th column equal to $b^{(k)}$. Algorithm 1 can be written equivalently as

$$B_h = \sum_{e=(p,c)\in E} \frac{h(p)-h(c)}{m_e} u_e u_e^\top. \quad (6)$$

When the height function is fixed, we use the abbreviation $B = B_h$. Define $\kappa(r) = 0$ at the root and, for every other node $v$,

$$\kappa(v) = \sum_{e\in P(r,v)} \frac{w_e}{m_e},$$

where $P(r, v)$ is the root-to-$v$ path. Then

$$B_{ij} = \kappa\big(\mathrm{lca}(\ell_i, \ell_j)\big). \quad (7)$$

Thus $B$ is the Gram matrix of a hierarchy-induced strong kernel (Kriege et al., 2016). An associated edge-indexed feature map is given by

$$\varphi_e(\ell_i) = \sqrt{\frac{w_e}{m_e}}\,\mathbf{1}\{\ell_i \in S_e\}, \qquad B_{ij} = \langle \varphi(\ell_i), \varphi(\ell_j)\rangle.$$

The columns $b^{(i)} = Be_i$ are kernel sections evaluated on the leaves. The metric introduced below is not the canonical RKHS distance induced by this kernel; it is total variation between the transformed probability profiles.

### 3.4 Theoretical properties

Proofs are given in Appendix A.

**Proposition 1** (Simplex preservation)**.** *For every leaf $\ell_k$, the vector $b^{(k)}$ constructed by Algorithm 1 satisfies*

$$b_j^{(k)} \geq 0 \quad \textit{for all } j, \qquad \sum_{j=1}^{L} b_j^{(k)} = 1.$$

*Hence $b^{(k)} \in \Delta^{L-1}$.*

**Theorem 2** (Exact recovery of leaf-topic distances)**.** *The vectors constructed by Algorithm 1 satisfy*

$$d_{\mathrm{TV}}\big(b^{(i)}, b^{(j)}\big) = h\big(\mathrm{lca}(\ell_i, \ell_j)\big) = H_{ij}$$

*for all $i, j \in \{1, \dots, L\}$.*

**Proposition 3** (Uniqueness within the edge-cluster operator class)**.** *Assume that unary internal vertices have been contracted. For coefficients $\alpha_e \geq 0$, let*

$$B_\alpha = \sum_{e\in E} \alpha_e u_e u_e^\top.$$

*If every column $B_\alpha e_i$ belongs to $\Delta^{L-1}$ and*

$$d_{\mathrm{TV}}(B_\alpha e_i, B_\alpha e_j) = H_{ij} \qquad \textit{for all leaves } i, j,$$

*then $\alpha_e = w_e/m_e$ for every edge $e$. Hence the normalization in Eq.* (6) *is unique within this class. This statement does not assert uniqueness among all possible simplex embeddings.*

*Remark* 1 (Unary taxonomic levels). The reduction used in Proposition 3 is an identifiability device, not a requirement to discard semantically meaningful taxonomic levels. Suppose that a unary vertex $v$ lies between $p$ and $c$, and that both edges subtend the same terminal-leaf set $S$, with indicator vector $u$. For any strictly intermediate height $h(c) < h(v) < h(p)$,

$$\frac{h(p) - h(v)}{|S|} uu^\top + \frac{h(v) - h(c)}{|S|} uu^\top = \frac{h(p) - h(c)}{|S|} uu^\top.$$

Consequently, reinserting $v$ does not change $B_h$, any distance between terminal topics, or the induced metric on leaf-supported mixed profiles. Evidence from terminal pairs does not identify how the total height drop is allocated between the two unary edges. That allocation matters only when internal nodes themselves are treated as geometric points.

**Proposition 4** (Stochasticity, positive definiteness, and invertibility)**.** *Let $B \in \mathbb{R}^{L\times L}$ be the matrix whose k-th column is $b^{(k)}$. Then B is nonnegative, symmetric, doubly stochastic, and positive definite. Consequently, B is invertible and its columns form a basis of $\mathbb{R}^L$.*

### 3.5 Worked example

Consider the seven-node tree in Fig. 1, four of whose nodes are leaves. The tree and its node heights determine the pairwise leaf distances shown in Table 1.

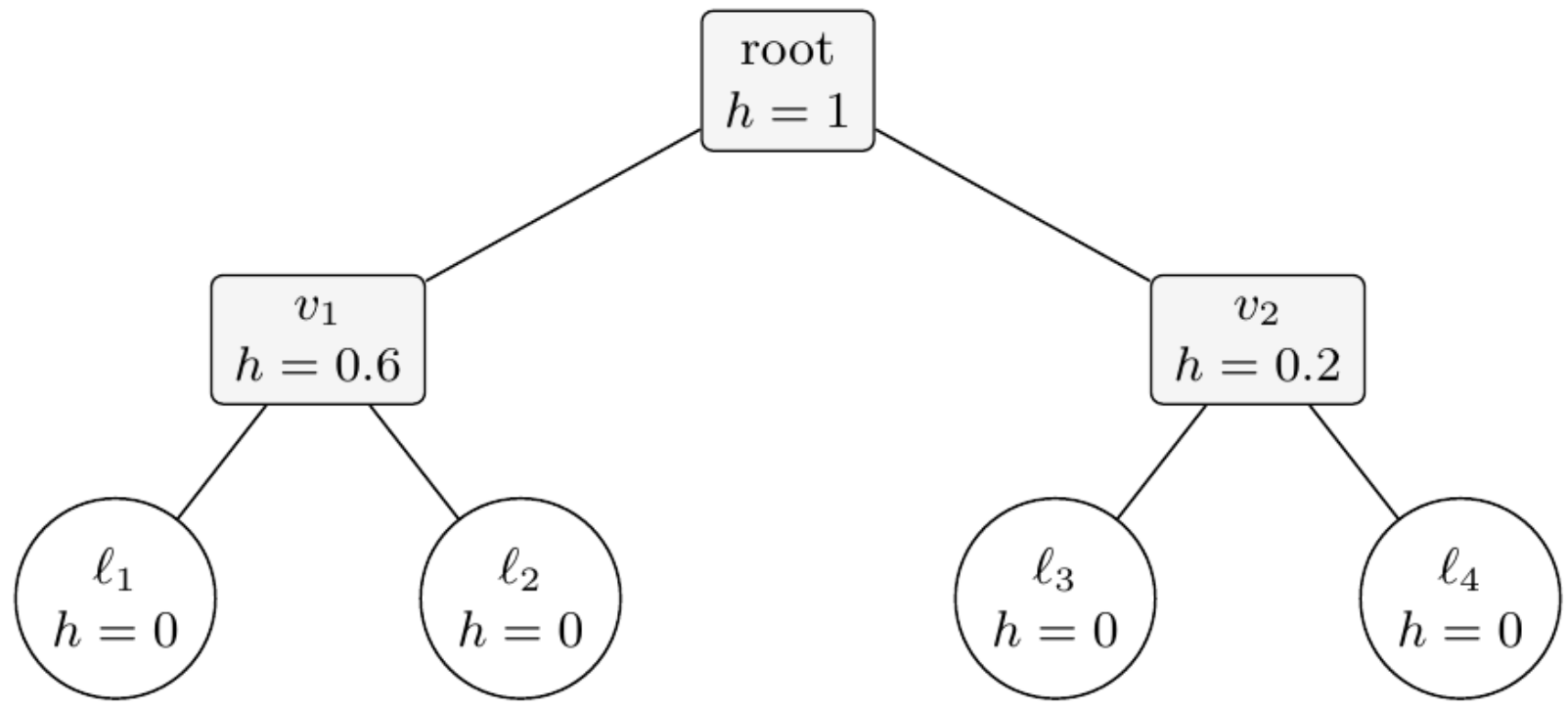

Figure 1: Example of a rooted tree with specified node heights

Algorithm 1 yields

$$b^{(1)} = (0.8, 0.2, 0, 0), \quad b^{(2)} = (0.2, 0.8, 0, 0),$$
$$b^{(3)} = (0, 0, 0.6, 0.4), \quad b^{(4)} = (0, 0, 0.4, 0.6).$$

Each vector lies in the probability simplex; together, the four vectors are linearly independent and realize the pairwise total variation distances in Table 1.

Table 1: Pairwise distances determined by the tree in Fig. 1

| $H_{ij}$ | $\ell_1$ | $\ell_2$ | $\ell_3$ | $\ell_4$ |
|---|---|---|---|---|
| $\ell_1$ | 0 | 0.6 | 1 | 1 |
| $\ell_2$ | 0.6 | 0 | 1 | 1 |
| $\ell_3$ | 1 | 1 | 0 | 0.2 |
| $\ell_4$ | 1 | 1 | 0.2 | 0 |

### 3.6 Metric for mixed topic profiles

Let $x = (x_1, \ldots, x_L) \in \Delta^{L-1}$ be the original topic profile of a scholarly entity. The hierarchy-adjusted image of $x$ in the same leaf-indexed coordinate system is

$$Bx = \left( \sum_{k=1}^{L} x_k b_1^{(k)}, \ldots, \sum_{k=1}^{L} x_k b_L^{(k)} \right)^{\mathsf{T}}.$$

Because the columns of $B$ are probability vectors, $Bx \in \Delta^{L-1}$. The original profile $x$ contains the observed Topic shares, whereas $B_h x$ is used only to compute distances. Because $B$ is invertible, $x$ can be recovered exactly from $B_h x$. For profiles $x, y \in \Delta^{L-1}$, define

$$D(x,y) = \frac{1}{2} \sum_{j=1}^{L} \left| \sum_{k=1}^{L} b_j^{(k)} (x_k - y_k) \right| = \frac{1}{2} \|B(x-y)\|_1. \tag{8}$$

When $B$ is the identity, Eq. (8) reduces to flat total variation. The metric $D$ is one particular linear extension of the prescribed pure-topic distances; the leaf ultrametric alone does not determine a unique metric on mixtures. It is a metric on the profile simplex and a pseudometric on scholarly entities whenever distinct entities may share the same profile.

**Corollary 5** (Metricity and contraction)**.** *The function $D$ defined in Eq.* (8) *is a metric on $\Delta^{L-1}$. Moreover,*

$$\frac{1}{\|B^{-1}\|_{1\to 1}} d_{\mathrm{TV}}(x,y) \le D(x,y) \le d_{\mathrm{TV}}(x,y) \qquad (x, y \in \Delta^{L-1}). \tag{9}$$

*Proof.* Nonnegativity and symmetry follow from the $\ell_1$ norm. Proposition 4 implies that $B$ is invertible, so $D(x,y) = 0$ if and only if $x = y$. Finally,

$$\begin{aligned} D(x,y) &= \frac{1}{2} \left\| B\left[(x-z) + (z-y)\right] \right\|_1 \\ &\le \frac{1}{2}\|B(x-z)\|_1 + \frac{1}{2}\|B(z-y)\|_1 \\ &= D(x,z) + D(z,y), \end{aligned}$$

which establishes the triangle inequality.

For a matrix $A$, write $\|A\|_{1\to 1} = \max_k \sum_j |A_{jk}|$ for the induced $\ell_1$ operator norm. Because $B$ is nonnegative and each column sums to one, $\|B\|_{1\to 1} = 1$. Hence

$$2D(x,y) = \|B(x-y)\|_1 \le \|B\|_{1\to 1} \|x-y\|_1 = 2d_{\mathrm{TV}}(x,y).$$

Conversely,

$$2d_{\mathrm{TV}}(x,y) = \|x-y\|_1 \le \|B^{-1}\|_{1\to 1} \|B(x-y)\|_1 = 2\|B^{-1}\|_{1\to 1} D(x,y),$$

which proves Eq. (9). □

Like flat total variation, $D$ permits a coordinate-level interpretation of overlap. Let $\tilde{x} = Bx$ and $\tilde{y} = By$. Coordinates $j$ for which $\min(\tilde{x}_j, \tilde{y}_j)$ is comparatively large contribute most strongly to the transformed overlap. Because the transformation redistributes mass among related leaves, these coordinates should be interpreted relative to the taxonomy and the matrix $B$, rather than as unchanged flat-topic weights.

Proposition 4 gives an additional interpretation. Under the column convention $B_{jk} = P(j \mid k)$, the matrix $B$ is a symmetric doubly stochastic Markov operator, and the uniform distribution is stationary. It redistributes each original profile toward hierarchically related leaves, which explains the contraction in Eq. (9). Invertibility ensures that this transformation does not identify distinct profiles.

### 3.7 Matrix-free evaluation

The dense matrix $B$ in Eq. (6) is not needed to evaluate a single distance. Put $z = x - y$ and, for every edge $e$, compute the signed subtree mass

$$s_e = \sum_{\ell_k \in S_e} z_k. \tag{10}$$

A bottom-up traversal computes all $s_e$. Equation (6) then gives

$$[Bz]_j = \sum_{e:\ell_j \in S_e} \frac{w_e}{m_e} s_e. \tag{11}$$

The right-hand side can be evaluated for all leaves in a single top-down traversal: when edge $e = (p, c)$ is crossed, add $(w_e/m_e)s_e$ to the value inherited from $p$. Finally,

$$D(x, y) = \frac{1}{2} \sum_{j=1}^{L} |[Bz]_j|.$$

Thus each pair of dense input profiles can be compared in $O(|V| + L)$ time using $O(|V| + L)$ working memory, once the edge weights and subtree sizes have been stored. This removes the $O(L^2)$ dense-matrix storage bottleneck for large topic taxonomies.

### 3.8 Relationship to tree-Wasserstein distance

The proposed metric and tree-Wasserstein distance use the same edge-height differences to recover the prescribed pure-leaf geometry, but aggregate mixed-profile mass differences differently. Let $x$ and $y$ be probability measures supported on the leaf set. Assign every edge $e$ the transport length $q_e = w_e/2$, so that the path length between two leaves equals their prescribed distance $H_{ij}$. For $\delta = x - y$, the standard tree formula is (Le et al., 2019; Peyré and Cuturi, 2019)

$$W_T(x, y) = \sum_{e \in E} q_e |s_e| = \frac{1}{2} \sum_{e \in E} w_e |s_e|, \qquad q_e = \frac{w_e}{2}. \tag{12}$$

**Proposition 6** (Comparison with tree-Wasserstein distance)**.** *For all $x, y \in \Delta^{L-1}$,*

$$D(x, y) \le W_T(x, y).$$

*The two distances agree whenever $x$ and $y$ are pure leaf profiles, but they need not agree for mixed profiles.*

*Proof.* Using Eq. (11) and the triangle inequality,

$$D(x,y) = \frac{1}{2}\sum_{j=1}^{L}\left|\sum_{e:\ell_j\in S_e}\frac{w_e}{m_e}s_e\right|$$
$$\le \frac{1}{2}\sum_{j=1}^{L}\sum_{e:\ell_j\in S_e}\frac{w_e}{m_e}|s_e| = \frac{1}{2}\sum_{e\in E} w_e|s_e| = W_T(x,y).$$

For pure profiles, Theorem 2 and the choice $q_e = w_e/2$ give the same leaf-to-leaf path distance $H_{ij}$. Strict inequality is demonstrated below. □

To illustrate strict inequality with three leaves, let the root $r$, with $h(r) = 1$, have children $a$ and $\ell_3$; let $h(a) = 1/2$, and let $a$ have leaf children $\ell_1, \ell_2$. Then

$$B = \begin{pmatrix} 3/4 & 1/4 & 0 \\ 1/4 & 3/4 & 0 \\ 0 & 0 & 1 \end{pmatrix}.$$

For $x = (3/10, 7/10, 0)$ and $y = (7/10, 0, 3/10)$, direct calculation gives $D(x,y) = 17/40$, whereas Eq. (12) gives $W_T(x,y) = 1/2$. The difference arises because signed contributions from nested clusters can cancel within a transformed coordinate before the absolute value is taken; tree-Wasserstein takes absolute values at the edge level. Thus the proposed distance is not a reparameterization of tree-Wasserstein.

Table 2 summarizes the differences among the flat baseline, the proposed operator, and tree-Wasserstein distance.

# 4 OpenAlex case study: text-assisted height initialization and empirical diagnostics

The construction in Section 3 requires an admissible height function but does not estimate it. Using a frozen OpenAlex taxonomy snapshot, we evaluated three aspects of the text-assisted initialization procedure: taxonomy–text alignment before correction, the monotonic adjustment needed to obtain admissible heights, and diagnostics of numerical recovery and sensitivity. We then illustrate how the calibrated distances apply to terminal Topics, internal categories, and mixed profiles.

## 4.1 Taxonomy snapshot and calibration protocol

The snapshot was retrieved from the OpenAlex Topics endpoint on 13 August 2026 and contained 4 Domains, 26 Fields, 252 Subfields, and 4,516 terminal Topics (Priem et al., 2022; OpenAlex, n.d.). With an artificial common root, the full taxonomy comprised 4,799 nodes, of which 283 were internal. The terminal Topics define the coordinates of the profile simplex in Section 2.

Thirty Subfields had only one Topic child, so cross-child comparisons cannot identify their heights. We therefore estimated text-based heights using the reduced branching tree containing 253 internal nodes and 4,516 terminal Topics, while retaining a mapping to the complete source taxonomy. By Remark 1, suppressing and reinserting a unary node leaves $B_h$ and all distances between leaf-supported profiles unchanged; only its height, or equivalently the allocation of the total height drop along that internal path, remains unidentified.

Each terminal Topic was represented using the frozen input template

```
{label} [SEP] {description}.
```

The primary encoder was the pinned 768-dimensional checkpoint

Table 2: Comparison of profile-distance constructions

| Property | Flat total variation | Proposed $D$ | Tree-Wasserstein $W_T$ |
|---|---|---|---|
| Taxonomy-aware | No | Yes | Yes |
| Pure-leaf geometry | Unit discrete distance | Prescribed ultrametric $H_{ij}$ | The same $H_{ij}$ represented as a path cost with $q_e = w_e/2$ |
| Mixed-profile construction | Direct coordinate comparison | Total variation after an invertible simplex-preserving operator | Optimal transport on tree edges |
| Representation used in distance evaluation | Original leaf-indexed profile | Transformed leaf-indexed profile $Bx$ | Signed subtree masses on tree edges |
| Tree-based evaluation | $O(L)$ | $O(\|V\| + L)$ | $O(\|V\| + L)$ |

`sentence-transformers/paraphrase-mpnet-base-v2`
revision `6cc9279c672dc57f94445ef259b28a1b736fec8f`.

The Sentence-Transformer framework is described by Reimers and Gurevych (2019). For terminal Topics $\ell_i$ and $\ell_j$, we used cosine dissimilarity

$$\delta_{ij} = 1 - \cos(\mathbf{e}_i, \mathbf{e}_j). \tag{13}$$

We use the term *dissimilarity* because Eq. (13) need not satisfy the triangle inequality.

For a branching node $v$ with child set $C(v)$, its raw height proposal under complete linkage was

$$\tilde{h}(v) = \max_{\substack{c_a, c_b \in C(v) \\ a<b}} \max_{\substack{\ell_i \in \mathcal{L}(c_a) \\ \ell_j \in \mathcal{L}(c_b)}} \delta_{ij}, \qquad \tilde{h}(\ell) = 0. \tag{14}$$

This is a conservative node-level summary: it is the smallest local upper bound on all raw dissimilarities between Topics that belong to different child branches of $v$. It should not be read as the raw dissimilarity of every pair whose lowest common ancestor is $v$.

Because independently computed proposals need not respect the parent–child order, we first computed the pointwise minimal non-strict monotone majorant and then enforced strict monotonicity. With both recursions initialized at zero on the leaves,

$$g_0(v) = \max\left\{\tilde{h}(v), \max_{c \in C(v)} g_0(c)\right\}, \tag{15}$$

$$g_\varepsilon(v) = \max\left\{\tilde{h}(v), \max_{c \in C(v)} \left[g_\varepsilon(c) + \varepsilon_0\right]\right\}, \qquad \varepsilon_0 = 10^{-6}. \tag{16}$$

Equation (15) is the pointwise smallest function that majorizes the raw proposals and is nondecreasing from child to parent in the original dissimilarity scale. Equation (16) additionally guarantees a positive height gap on every edge. The reported text-initialized heights were then normalized according to

$$h(v) = \frac{g_\varepsilon(v)}{g_\varepsilon(r)}, \tag{17}$$

where $r$ is the artificial root. Normalization preserves strict node ordering; the least-majorant property applies to $g_0$ on the original dissimilarity scale, before normalization.

Substituting Eq. (17) into Eq. (6) gives

$$B_h = \frac{1}{g_\varepsilon(r)} \sum_{e=(p,c) \in E} \frac{g_\varepsilon(p) - g_\varepsilon(c)}{m_e} u_e u_e^\top. \tag{18}$$

Thus the simplex operator is built from the normalized height function rather than directly from the unnormalized envelope.

### 4.2 Intrinsic taxonomy–text alignment before correction

We evaluated the raw embeddings before applying Eqs. (15)–(17) or constructing $B$. For each anchor Topic, candidate comparators were partitioned into four ordered categories: $C_0$, Topics in the same Subfield as the anchor; $C_1$, Topics in the same Field but a different Subfield; $C_2$, Topics in the same Domain but a different Field; and $C_3$, Topics in a different Domain. For each adjacent contrast, 32 nearer–farther pairs were sampled with replacement for every eligible anchor. Let $\mathcal{I}_k$ be the set of anchors for which both categories are nonempty, let $R = 32$, and let $J_{ik}^{(r)}$ and $J_{i,k+1}^{(r)}$ be independent uniform draws from $C_k(i)$ and $C_{k+1}(i)$. The anchor-balanced estimate was

$$\widehat{A}_k = \frac{1}{|\mathcal{I}_k|R} \sum_{i\in\mathcal{I}_k} \sum_{r=1}^{R} \left[ \mathbf{1}\left\{\delta_{iJ_{ik}^{(r)}} < \delta_{iJ_{i,k+1}^{(r)}}\right\} + \frac{1}{2}\mathbf{1}\left\{\delta_{iJ_{ik}^{(r)}} = \delta_{iJ_{i,k+1}^{(r)}}\right\}\right]. \tag{19}$$

Anchors, rather than all available Topic pairs, receive equal weight; large Subfields or Fields therefore cannot dominate the estimand merely by containing more Topics. A value of 0.5 indicates no ordering, whereas values above 0.5 indicate that the comparator nearer in the taxonomy also tends to have lower raw text dissimilarity. All three ordering scores exceeded 0.69 (Table 3). The first contrast excluded the 30 Topics in singleton Subfields because $C_0$ was empty for those anchors. The 95% intervals are stability intervals obtained by bootstrapping anchors for this fixed taxonomy (1,000 resamples); no exact ties occurred in the sampled comparisons.

To assess alignment simultaneously across all three levels, we used the weakest-link statistic

$$T_{\min} = \min(\widehat{A}_0, \widehat{A}_1, \widehat{A}_2) = 0.691361. \tag{20}$$

Embedding vectors were reassigned within ten strata defined by input-text length in characters, thereby breaking the association between embeddings and taxonomic positions while preserving the coarse length distribution. None of 1,999 stratified permutations produced a statistic at least as large as the observed value. The plus-one Monte Carlo estimate was therefore

$$\widehat{p}_{\mathrm{MC}} = \frac{0+1}{1999+1} = 0.0005, \tag{21}$$

with a 95% Clopper–Pearson interval of $[0, 0.001844]$ for the conditional permutation exceedance probability. Under the stratified permutation null hypothesis, this result provides evidence of ordinal taxonomy–text association.

### 4.3 Extent of monotonic correction, numerical recovery, and sensitivity

Every unordered pair of terminal Topics was assigned to its unique lowest common ancestor. Consequently, all

$$\binom{4516}{2} = 10\,194\,870 \tag{22}$$

pairs entered the cross-child calculations exactly once. This is complete computational coverage, not a sample of independent observations.

We decomposed the normalized upward correction into a taxonomic component and a strict-margin component:

$$C_{\mathrm{tax}}(v) = \frac{g_0(v) - \tilde{h}(v)}{g_\varepsilon(r)}, \qquad C_{\mathrm{margin}}(v) = \frac{g_\varepsilon(v) - g_0(v)}{g_\varepsilon(r)}. \tag{23}$$

Only four raw parent–child edges violated monotonicity, and the non-strict envelope consequently raised just 3 of the 253 calibrated internal nodes (1.19%). All three were Fields (Table 4); the median and 95th percentile of $C_{\mathrm{tax}}$ across internal nodes were both zero. The largest normalized lift was 0.143080 for Veterinary, making this node a priority for independent expert review of its calibration and semantic placement. The normalized margin introduced to enforce strict monotonicity was only $9.10 \times 10^{-7}$. All height columns in Table 4 use the common denominator $g_\varepsilon(r)$.

Table 3: Intrinsic ordering of raw OpenAlex Topic-text dissimilarities

| Nearer versus farther categories | Eligible anchors | Comparisons | $\widehat{A}_k$ | 95% stability interval |
|---|---|---|---|---|
| Same Subfield vs. another Subfield in the same Field | 4,486 | 143,552 | 0.692857 | [0.687451, 0.697852] |
| Another Subfield in the same Field vs. another Field in the same Domain | 4,516 | 144,512 | 0.691361 | [0.686551, 0.695977] |
| Another Field in the same Domain vs. another Domain | 4,516 | 144,512 | 0.701443 | [0.696917, 0.706215] |

Table 4: Internal nodes changed by the monotone envelope

| Field | Raw height | Non-strict height | Final strict height | $C_{\text{tax}}$ |
|---|---|---|---|---|
| Chemistry | 0.783742 | 0.804674 | 0.804674 | 0.020931 |
| Energy | 0.693139 | 0.717982 | 0.717983 | 0.024843 |
| Veterinary | 0.432716 | 0.575796 | 0.575797 | 0.143080 |

The resulting $4,516 \times 4,516$ operator satisfied the theoretical invariants to floating-point precision. The maximum column-stochasticity residual was $6.75 \times 10^{-14}$, and the maximum discrepancy between $\frac{1}{2}\|B(e_i - e_j)\|_1$ and $h(\text{lca}(\ell_i, \ell_j))$ on the deterministic LCA-stratified validation set was $3.33 \times 10^{-16}$.

Two sensitivity analyses varied one modeling choice at a time. The tail variant retained MPNet but replaced each cross-child block maximum by its empirical 0.95 quantile before taking the maximum across blocks. The encoder variant retained the maximum statistic but replaced MPNet with SPECTER2 (Singh et al., 2023). Table 5 shows that the tail variant largely preserved the rank order of heights, whereas changing the encoder produced a lower rank correlation and a substantially larger maximum node-level height change.

The pair-weighted mean differences understate local sensitivity because $71.73\%$ of Topic pairs meet at the root and therefore remain at distance 1 in every variant. Rank-specific and node-level changes are therefore more informative than the overall pair-weighted mean. With $\rho = 0.789$ and a maximum $|\Delta h|$ of 0.382, the primary height estimates depend materially on encoder choice.

### 4.4 Interpreting the resulting distances

The calibrated tree yields three distinct types of quantities, which should not be conflated. For terminal Topics—the coordinates of the simplex— the construction gives

$$D(e_i, e_j) = h\big(\text{lca}(\ell_i, \ell_j)\big). \tag{24}$$

To make Eq. (24) concrete, we selected three terminal Topics drawn from mathematics, physics, and psychology: Algebraic Geometry and Number Theory (`T10061`), Quantum Mechanics and Applications (`T10622`), and Cognitive Functions and Memory (`T13471`). Table 6 contrasts the raw pairwise dissimilarities with the corresponding distances induced by the calibrated tree.

Under the calibrated tree geometry, the two Topics in the Physical Sciences Domain are closer to each other than either is to the Topic in Psychology. Their tree-induced distance, 0.943557, is the calibrated height of their lowest common ancestor rather than their raw pairwise dissimilarity; the three raw values themselves do not form a tree metric.

"Mathematics," "Physics and Astronomy," and "Psychology" are Field-level internal nodes rather than leaf coordinates. Write $M$, $F$, and $P$ for these three Fields, respectively. If internal categories are treated as vertices of the tree with edge lengths $q_e = w_e/2$, their path distance is

$$d_q(u, v) = h\big(\text{lca}(u, v)\big) - \frac{h(u) + h(v)}{2}. \tag{25}$$

Table 5: Sensitivity of the text-initialized geometry

| Variation | Spearman $\rho$ for heights | Max. $\lvert\Delta h\rvert$ | Pair-weighted mean $\lvert\Delta H\rvert$ | Lifted nodes |
|---|---|---|---|---|
| Within-block 0.95 quantile | 0.992846 | 0.098509 | 0.011648 | 21 |
| SPECTER2 encoder | 0.789425 | 0.382350 | 0.010138 | 3 |

Table 6: Raw and calibrated distances for three illustrative terminal OpenAlex Topics

| Topic pair | Lowest common ancestor | Raw $\delta_{ij}$ | Tree-induced $D$ |
|---|---|---|---|
| Algebraic Geometry and Number Theory – Quantum Mechanics and Applications | Physical Sciences | 0.523046 | 0.943557 |
| Algebraic Geometry and Number Theory – Cognitive Functions and Memory | Root | 0.750062 | 1.000000 |
| Quantum Mechanics and Applications – Cognitive Functions and Memory | Root | 0.675857 | 1.000000 |

The fitted heights are $h(M) = 0.786385$, $h(F) = 0.818626$, and $h(P) = 0.775545$; Physical Sciences has height 0.943557 and the root has height 1. Hence

$$d_q(M,F) = 0.943557 - \frac{0.786385 + 0.818626}{2} = 0.141051,$$
$$d_q(M,P) = 1 - \frac{0.786385 + 0.775545}{2} = 0.219035,$$
$$d_q(F,P) = 1 - \frac{0.818626 + 0.775545}{2} = 0.202914.$$

The Fields ”Mathematics” and ”Physics and Astronomy” are therefore the closest pair among these three internal categories under $d_q$. These values are neither distances between publication profiles nor distances between Field centroids in the leaf simplex. For mixed profiles, $D$ is computed from the geometric image $Bx$. Descriptive summaries at an OpenAlex rank $R$ should instead aggregate the original profile $x$ as

$$x_v^{(R)} = \sum_{\ell_j \in \mathcal{L}(v)} x_j. \tag{26}$$

This rank aggregate is a profile summary; it does not represent the internal vertex $v$ as a geometric point.

The following three-component example shows how distances among the components affect mixed-profile distances. Let $M$, $F$, and $P$ be three illustrative components satisfying $H_{MF} = 0.25$ and $H_{MP} = H_{FP} = 1$, giving

$$B = \begin{pmatrix} 0.625 & 0.375 & 0 \\ 0.375 & 0.625 & 0 \\ 0 & 0 & 1 \end{pmatrix}.$$

For

$$x_A = (0.8, 0.1, 0.1), \qquad x_B = (0.2, 0.7, 0.1), \qquad x_C = (0.2, 0.1, 0.7),$$

flat total variation gives 0.6 for every pair, whereas Eq. (8) gives

$$D(x_A, x_B) = 0.15, \qquad D(x_A, x_C) = D(x_B, x_C) = 0.60.$$

Each pair of profiles differs by the same amount of mass, but in the first pair the difference lies between nearby components, whereas in the other two the difference crosses the root split. The value 0.25 is illustrative rather than estimated from OpenAlex; the example demonstrates the behavior of the metric, not its empirical utility.

# 5 Discussion

The study separates two problems that should be evaluated differently: realizing a supplied hierarchical geometry on the probability simplex, and estimating the height function that specifies that geometry. The first problem is solved exactly under the assumptions of Section 3; the OpenAlex analysis addresses the second by testing a text-assisted initialization procedure.

## 5.1 Scientometric interpretation and applications

In scientometric applications, profile estimation and distance calibration remain distinct. The observed profile $x$ can be obtained from any compatible topic-assignment procedure, while the taxonomy and its height function determine how differences in Topic shares across coordinates contribute to the resulting distance. Changes in the height calibration therefore alter the comparison geometry without redefining the reported Topic shares. The OpenAlex results suggest that Topic texts exhibit sufficient ordinal alignment with the taxonomy to initialize this geometry, but the encoder sensitivity shows that the numerical heights remain a modeling choice rather than a fixed property of the taxonomy.

Potential applications of the metric include publication retrieval, reviewer recommendation, portfolio comparison, and science mapping. In each application, the taxonomy and calibrated heights remain explicit model inputs, while the matrix-free algorithm permits repeated comparisons without storing a dense $L \times L$ matrix.

## 5.2 Limitations and validation requirements

Several limitations remain. The model uses only the rooted-tree backbone of an ontology and therefore omits cross-cutting semantic relations. Its geometry also depends on Topic assignment, taxonomy version and granularity, text fields, encoder, linkage statistic, and normalization. Because the OpenAlex initialization uses Topic texts supplied by the same system that defines the taxonomy, the resulting calibration still requires external validation against expert similarity judgments. The study does not establish that $D$ improves retrieval, recommendation, portfolio comparison, or science mapping. External validation should evaluate $D$, flat total variation, cosine- and Jensen–Shannon-based alternatives, similarity-adapted publication vectors (Rousseau et al., 2017), and tree-Wasserstein distance against expert similarity judgments and on preregistered retrieval or ranking tasks.

# 6 Conclusions

Flat total variation ignores the hierarchical relations among topic coordinates. We showed that, for a rooted taxonomy with an admissible height function, a cardinality-normalized operator $B_h$ realizes the corresponding lowest-common-ancestor ultrametric exactly within the same probability simplex. The same invertible transformation defines a hierarchy-aware metric on mixed topic profiles and admits $O(|V| + L)$ matrix-free evaluation.

The OpenAlex case study demonstrates a practical, though model-dependent, procedure for text-assisted initialization of the height function. Accordingly, the observed topic profile remains directly interpretable, whereas calibrated heights define the comparison geometry and serve as starting values for expert calibration.

# A Proofs of the theoretical results

*Proof of Proposition 1.* By construction,

$$b_j^{(k)} = \sum_{\substack{e=(p,c)\in E:\\ \ell_k,\ell_j\in S_e}} \frac{h(p) - h(c)}{|S_e|}.$$

Every term is nonnegative because $h$ decreases along each edge. Hence $b_j^{(k)} \geq 0$ for all $j, k$.

Fix $\ell_k$ and sum over the coordinates:

$$\sum_{j=1}^{L} b_j^{(k)} = \sum_{\substack{e=(p,c)\in E:\\ \ell_k \in S_e}} \sum_{\ell_j \in S_e} \frac{h(p) - h(c)}{|S_e|}$$
$$= \sum_{\substack{e=(p,c)\in E:\\ \ell_k \in S_e}} \big(h(p) - h(c)\big).$$

The relevant edges are precisely those on the path from the root to $\ell_k$. The sum telescopes to

$$h(\text{root}) - h(\ell_k) = 1.$$

Thus $b^{(k)} \in \Delta^{L-1}$. □

*Proof of Theorem 2.* The case $i = j$ is immediate. Assume henceforth that $i \neq j$, fix $\ell_i, \ell_j$, and let $a = \mathrm{lca}(\ell_i, \ell_j)$. Write $E_i$ and $E_j$ for the edge sets on the paths from $a$ to $\ell_i$ and $\ell_j$, respectively. Contributions from edges for which $S_e$ contains either both leaves or neither leaf cancel in $b^{(i)} - b^{(j)}$. Hence

$$b^{(i)} - b^{(j)} = \sum_{e\in E_i} \frac{w_e}{m_e} u_e - \sum_{e\in E_j} \frac{w_e}{m_e} u_e.$$

All terms in the first sum are nonnegative and supported in the child subtree of $a$ containing $\ell_i$; all terms in the second sum have the opposite sign and are supported in the disjoint child subtree containing $\ell_j$. There is therefore no cancellation inside the $\ell_1$ norm. Since $\|u_e\|_1 = m_e$,

$$d_{\mathrm{TV}}\big(b^{(i)}, b^{(j)}\big) = \frac{1}{2}\left(\sum_{e\in E_i} w_e + \sum_{e\in E_j} w_e\right) = \frac{1}{2} \sum_{e\in P(\ell_i,\ell_j)} w_e.$$

Telescoping along the two branches gives

$$\frac{1}{2} \sum_{e\in P(\ell_i,\ell_j)} w_e = \frac{1}{2}\big[h(a) - h(\ell_i) + h(a) - h(\ell_j)\big]$$
$$= \frac{1}{2}[h(a) + h(a)]$$
$$= h(a) = H_{ij}.$$

This proves Eq. (5). □

*Proof of Proposition 3.* Put $\beta_e = \alpha_e m_e$. Because $\alpha_e \geq 0$, the entries of $B_\alpha$ are nonnegative. The requirement that the column $B_\alpha e_i$ lie in the simplex therefore gives, for every leaf $\ell_i$,

$$\sum_{e\in P(r,\ell_i)} \beta_e = 1. \tag{A.1}$$

For a node $v$, define $t(v)$ as the $\beta$-weight of a path from $v$ to any descendant leaf. This value is well defined: by Eq. (A.1), it equals one minus the common $\beta$-weight of the path from the root to $v$, independently of the chosen descendant leaf. In particular, $t(r) = 1$ and $t(\ell) = 0$ for every leaf.

Every internal vertex of the reduced tree has at least two children. Choose leaves $\ell_i$ and $\ell_j$ in two distinct child subtrees of an internal vertex $v$; then $v = \mathrm{lca}(\ell_i, \ell_j)$. As in the proof of Theorem 2, the two branch contributions to $B_\alpha e_i - B_\alpha e_j$ have disjoint supports and opposite signs. Therefore

$$d_{\mathrm{TV}}(B_\alpha e_i, B_\alpha e_j) = \frac{1}{2}\left(\sum_{e\in P(v,\ell_i)} \beta_e + \sum_{e\in P(v,\ell_j)} \beta_e\right) = t(v).$$

Exact recovery gives $t(v) = H_{ij} = h(v)$. The same identity already holds at the leaves, where both sides are zero. Hence, for every edge $e = (p, c)$,

$$\beta_e = t(p) - t(c) = h(p) - h(c) = w_e.$$

It follows that $\alpha_e = \beta_e / m_e = w_e / m_e$ for all $e \in E$. □

*Proof of Proposition 4.* For each edge $e = (p, c)$, define the indicator vector $u_e \in \{0, 1\}^L$ by

$$(u_e)_k = \begin{cases} 1, & \ell_k \in S_e, \\ 0, & \text{otherwise.} \end{cases}$$

The vector associated with leaf $\ell_k$ is

$$b^{(k)} = \sum_{e:\ell_k \in S_e} \frac{w_e}{m_e} u_e. \tag{A.2}$$

Therefore the matrix whose columns are these vectors has the decomposition in Eq. (6). Every coefficient and every entry of each outer product is nonnegative, so $B$ is nonnegative. Each summand is symmetric, so $B$ is symmetric. Proposition 1 shows that every column sums to one; symmetry then shows that every row also sums to one. Hence $B$ is doubly stochastic.

For any nonzero $a = (a_1, \dots, a_L)^\top$,

$$\begin{aligned} a^\top B a &= \sum_{e \in E} \frac{w_e}{m_e} (a^\top u_e)^2 \\ &= \sum_{e \in E} \frac{w_e}{m_e} \left( \sum_{\ell_k \in S_e} a_k \right)^2. \end{aligned}$$

Every term is nonnegative because $w_e > 0$. Equality can hold only if, for every edge $e$,

$$\sum_{\ell_k \in S_e} a_k = 0. \tag{A.3}$$

For each leaf $\ell_j$, the edge joining its parent to $\ell_j$ has the singleton cluster $S_e = \{\ell_j\}$. Equation (A.3) then gives $a_j = 0$. This holds for all $j$, contradicting $a \neq 0$. Hence $B$ is positive definite and therefore invertible. Its $L$ columns are linearly independent and form a basis of $\mathbb{R}^L$. □